\documentstyle[twocolumn,prl,aps,epsf]{revtex}

\begin{document}

\draft

\twocolumn[
\hsize\textwidth\columnwidth\hsize\csname@twocolumnfalse\endcsname

\title{Preon Trinity}
\author{Jean-Jacques Dugne,$^1$
Sverker Fredriksson,$^2$ Johan Hansson,$^2$
and Enrico Predazzi$^3$}
\address{
$^1$Laboratoire de Physique Corpusculaire, \\
Universit\'{e} Blaise Pascal, Clermont-Ferrand II,
FR-63177  Aubi\`{e}re, France \\
$^2$Department of Physics, Lule\aa \ University of Technology,
SE-97187 Lule\aa , Sweden \\
$^3$Department of Theoretical Physics,
Universit\'{a} di Torino,
IT-10125 Torino, Italy}


\maketitle

\begin{abstract}
We present a new minimal model
for the substructure of all known
quarks, leptons and weak gauge bosons,
based on only three fundamental and stable
spin-$1/2$ preons. As a consequence, we predict
three new quarks, three new leptons,
and six new vector bosons. One of
the new quarks has charge $-4e/3$.
The model explains the apparent
conservation of three lepton numbers,
as well as the so-called Cabibbo-mixing
of the $d$ and $s$ quarks, and predicts
electromagnetic decays or oscillations
between the neutrinos $\bar{\nu}_{\mu}$
($\nu_{\mu}$) and $\nu_e$ ($\bar{\nu}_e$).
Other neutrino oscillations, as well as
rarer quark mixings and CP violation
can come about due to a small
quantum-mechanical mixing of two of
the preons in the quark and lepton
wave functions.

\end{abstract}

\pacs{PACS numbers: 12.60.Rc, 13.35.-r, 14.60.-z, 14.65.-q}

]

\narrowtext

The arguments in favor of a
substructure of quarks and leptons in terms
of more fundamental preons have been
discussed for two decades (a review is
given in \cite{preons}). The most
compelling ones are that there exist
so many quarks and leptons, and that
they seemingly form a pattern of
three ``families''.

Much work in the field has been
inspired also by the fact that the
weak gauge bosons, $Z^0$ and
$W^{\pm}$, can be understood
as being composite. Hence,
the weak force would not be fundamental,
but rather a ``leakage'' of  multi-preon
states, in analogy to the
nuclear force being mediated by
quark-antiquark states. If so, there is no
fundamental electroweak unification,
and no theoretical complications
caused by massive, unstable
and electrically charged gauge bosons.
Consequently, there is no need for the so-called
Higgs mechanism with its many by-products,
{\it e.g.}, ``multiple Higgs'',
``composite Higgs'', and ``higgsinos'',
motivated mainly by internal theoretical
problems \cite{higgs}.

Another hint at
a substructure is that most quarks and
leptons are {\it unstable},
which in our view disqualifies them as fundamental
particles. Historically, all decays of
``elementary'' particles have sooner
or later been interpreted in terms
of processes among
even more fundamental objects.
This wisdom has not yet been
included in the so-called standard model,
where all quarks
and leptons are considered
fundamental, stable or not.

Preon models have so far focused on explaining
the lightest family of two quarks ($u$ and $d$)
and two leptons ($e$ and $\nu_e$)
with as few preons as possible. This means either
a minimal number of (two) different
preons, {\it e.g.}, the ``rishons''
\cite{harari,shupe}, or a minimal number
of (two) preons inside a quark or lepton,
{\it e.g.}, the ``haplons'' \cite{fritzsch}.

The two heavier families are usually
considered ``excitations''
of the light one, but no simple, consistent
and predictive model exists.
Neither have preon models
(or the standard model) been
able to explain the
conservation of three different lepton
numbers, the mixing of quarks, but not
of leptons, as described by the
Cabibbo-Kobayashi-Maskawa formalism
\cite{CKM} (quantified by the
so-called CKM matrix), and the CP violation
in neutral $K$ meson decays.

A classical problem with all preon models
(including ours) is that neutrinos are
so much lighter than the
quarks, especially as preons are expected
to have masses in the TeV range.
This is related to a lack
of understanding of the
forces that keep quarks and leptons together,
and also to the problem of defining a mass
of a permanently bound object.

The essence of our new preon model is that
we explain {\it all} known
quarks and leptons with the help of a
minimal number of
preons, while maintaining as much
symmetry and simplicity as
possible. We have been inspired by the haplon model
of Fritzsch and Mandelbaum \cite{fritzsch},
where preons have spin $1/2$ and $0$.
In addition, we guess
that there is a pairwise (``supersymmetric'')
relation between preons of different spin
as far as their charges and
composite states are concerned.

It turns out that one needs only
three preons of each spin to build
all known leptons and quarks.
We call the spin-$1/2$ preons $\alpha$,
$\beta$ and $\delta$, and the
spin-$0$ ones $x$, $y$ and $z$.
There are several reasons to believe
that the scalar preons (``spreons'') are indeed
composite themselves,
being spin-$0$ bound states (``di-preons'') of
two spin-$1/2$ preons in a symmetric way.
The notion of a spreon, which refers to
names like ``squark'' and
``slepton'' within the established theory
of supersymmetry (SUSY),
is only to remind about a
purely phenomenological symmetry
in our model.
It is similar to the ``supersymmetry''
between quarks and diquarks,
as sometimes quoted in quark-diquark
models of baryons \cite{anselm}.

It is worth noting that even if
the spreons were truly fundamental,
and SUSY partners of
preons, the symmetry would {\it not} be
implemented among quarks
and leptons. Hence, ``squarks'' and ``sleptons''
need not exist, and the real SUSY
partners of quarks and leptons would
themselves be quarks and leptons!

Our full preon scheme, containing
three fundamental spin-$1/2$ particles,
reads:

\vspace{3mm}

\begin{tabular}{l|c|c|c|}
charge & $+e/3$ & $-2e/3$ & $+e/3$ \\ \hline
spin-$1/2$ preons & $\alpha$ & $\beta$ & $\delta$ \\
spin-$0$ ``spreons'' & $x = (\bar{\beta} \bar{\delta})$
& $y = (\bar{\alpha} \bar{\delta})$
& $z = (\bar{\alpha} \bar{\beta})$ \\
\end{tabular}

\vspace{3mm}

We assume that $\alpha$, $\beta$ and $\delta$
carry a quantum chromodynamic (QCD)
color charge ({\it anti}color), {\it i.e.},
the preon colors are the $\bf 3^*$
representation of $SU(3)_{color}$, when
the quark colors are $\bf 3$.
By construction, also the
$x$, $y$ and $z$ are then color-${\bf 3^*}$
($\bf 3 \otimes 3 = 3^* \oplus 6$).
With quarks and leptons as, respectively,
preon-spreon and preon-antispreon
states, the colour representations correctly become
{\bf 3} for quarks and {\bf 1} for leptons,
as they should. We assume that the other
mathematically allowed color
configurations do not exist in nature.

The preon assignments we propose are summarised
in the table below, where the result of
combining a preon with a spreon or an
antispreon can be found in the respective cells.

\vspace{3mm}

\begin{tabular}{l|c|c|c|c|c|c|} 
& $\bar{x}$
& $\bar{y}$
& $\bar{z} $
& $x$
& $y$
& $z$  \\
& $(\beta \delta)$
& $(\alpha \delta)$
& $(\alpha \beta)$
& $(\bar{\beta} \bar{\delta})$
& $(\bar{\alpha} \bar{\delta})$
& $(\bar{\alpha} \bar{\beta})$  \\ \hline
$\alpha$
& $\nu_e$
& $\mu^+$ & $\nu_{\tau}$
& $u$
& $s$ & $c$ \\
$\beta$ & $e^-$
& $\bar{\nu}_{\mu}$
& $\tau^-$
& $d$
& $X$ & $b$  \\
$\delta$
& $\bar{\nu}_{\kappa 1}$
& $\kappa^+$
& $\bar{\nu}_{\kappa 2}$
& $t$
& $g$ & $h$  \\
\end{tabular}

\vspace{3mm}

There are alternative schemes for
building the known quarks and leptons,
but these turn out to be less consistent
in details, and will not be discussed here.

In the remainder of this work, we discuss
several aspects of this model,
starting with the trivial observation that {\it all}
known quarks and leptons are reproduced
with their correct charges, color-charges and spins.

Our scheme does not have the
three-family structure, so much quoted,
but not understood, within the standard model.
There is virtually a pairwise
relation between quarks and leptons, but not
with the same pairs as conventionally
defined. The true mathematical structure
should instead be ``$SU(3)_{preon}$'',
{\it i.e.}, a preon version of the
flavor-$SU(3)$ adopted for the three
lightest quarks ($u$, $d$ and $s$),
although the symmetry
must be severely broken, as far as
quark and lepton masses are concerned.

As the preons are flavor-${\bf 3}$,
the three spreons are ${\bf 3}$ too.
Hence, the leptons
are $\bf 3 \otimes 3^* = 1 \oplus 8$,
and the quarks are
$\bf 3 \otimes 3 = 3^* \oplus 6$. From
the table above it is obvious that the
lepton flavor-singlet ${\bf 1}$
is to be found among the three neutrinos
($\nu_{e},\bar{\nu}_{\mu},\bar{\nu}_{\kappa 2}$),
along the diagonal. It is not
possible to pinpoint any particular one,
or a linear combination, as the singlet,
or make an exact analogy to the
baryon octet in the quark model,
because of the symmetry breaking.
It is less ambiguous that
the three diagonal quarks
($u$,$X$,$h$) are flavor-$\bf 3^*$, while the
remaining ones are flavor-$\bf 6$.

The scheme contains
{\it three new leptons} and {\it three
new quarks}, which could all be
too heavy to be created in
current experiments, but naturally
of interest for future accelerator
energies in the TeV range. Among both leptons
and quarks there are two clear mass-trends.
Firstly, the bottom row contains objects
that are much heavier than the others,
which must be caused by a ``naked''
superheavy $\delta$
preon. The presence of a $\delta$ does
not, however, make a
spreon superheavy, since
$(\alpha \beta)$ seems to be
heavier than $(\beta \delta)$ and
$(\alpha \delta)$.
Secondly, there is an
increasing mass along the
diagonals from upper left to lower right.

The $t$ quark is ``predicted'' to
be superheavy, which explains
why it is so much heavier
than its conventionally assigned partners,
the $b$ quark and the $\tau$ lepton.
These enormous mass
differences are hard to accept
as {\it ad hoc} consequences of the
Higgs mechanism in the standard model.

One of the predicted new quarks
is the $X$ of charge
$-4e/3$, which seemingly does not belong to the
superheavy group. This naturally poses
a problem to the model, and the apparent lack
of experimental indications
must be analysed.

First, we note that quarks belonging
to different flavor-$SU(3)$ representations
need not have related masses.
Neither would quark masses need to
be simply related to lepton masses.
Therefore, it might well happen that the
flavor-$\bf 3^*$ quarks have rapidly
increasing masses, from $u$ to 
$X$ and $h$. Then the quark
masses would depend not only
on the underlying preon masses,
but also strongly on the exact composition
of the internal wave functions. Ultimately,
all $SU(3)$ representations need not be realised
in nature. The rule of the game is,
however, that if one object exists, so should
the full representation, which makes it difficult
to claim that the $X$ ``need not exist''.

A second alternative is that the $X$ quark
might indeed be the one that is
commonly believed to be the top quark,
and that the conventional charge assignment
($+2e/3$) of the ``top'' is incorrect.
This heretic idea cannot
easily be dismissed, since there
is no absolute experimental evidence
that the top charge is really $+2e/3$
\cite{top}. The top quark was found
through its presumed main decay channels,
namely semi-leptonic ones like
$t \rightarrow b + \ell^+ + \nu$, or
non-leptonic ones like
$t \rightarrow b + u + \bar{d}$,
where a few $b$ quarks have
been ``tagged'' by a charged muon
in semi-leptonic decays.
However, the situation is rather
complex because a total event contains
the decay of the full $t \bar{t}$ pair
into several leptons and hadronic jets,
and the identification and matching of
those are non-trivial. In addition, there are
individual events that seemingly
contradict the assumption of top-quark
production. It can therefore not be excluded
that a produced ``$b$'' is, in fact, an $\bar{b}$
(and {\it vice versa}), or that a tagged $b$
has been matched with the wrong lepton,
{\it i.e.}, from a $W$ decay of the
other quark.
These are crucial questions, since the
corresponding decay channels for the
antiquark $\bar{X}$ would be
$\bar{X} \rightarrow \bar{b} + \ell^+ + \nu$
and $\bar{X} \rightarrow \bar{b} + u + \bar{d}$,
so that the full $X \bar{X}$ decay would
give the same leptons and jets
as in a $t \bar{t}$ decay, although with
a different $b-W$ matching.
This puzzle would require a full new
analysis of the ``top'' data, where all
matchings are remade with the assumption
of $X \bar{X}$ decay.

A third, and even more speculative
idea, is that a ``light'' $X$ quark
has escaped discovery. All searches for
a new quark seem to focus on
a ``fourth-family'' $b'$ with
charge $-e/3$ \cite{rpp}.
They also rely on model-dependent
assumptions on $b'$ decay,
{\it e.g.}, through a flavor-changing
neutral current in
$b' \rightarrow b + \gamma$ \cite{abachi}
(searched for with gamma detectors).
Such experiments cannot see a decay of
the $X$ as given above.
Neither has there been systematic searches
for new resonances in $e^+e^-$ collisions
at high energies in the traditional way of
fine-tuning the total energy in small steps.
It is, on the other hand, hard to believe
that a light $X$ would not have been detected
in the search for the top quark \cite{top},
unless there is some (narrow)
kinematic region that
has not been properly covered.
Also, an $X$ quark lighter than
the $W^-$ should
have been seen through decays like
$W^- \rightarrow X + \bar{d}$ in the
$W$ branching ratios \cite{rpp}.
The case for an $X$ with a mass below,
say, the $W$ mass is therefore weak.

Turning now to {\it lepton decays},
we note that all the known ones can be
described as regroupings of preons into less
massive states. A decaying
charged lepton transfers its spreon to
a neutrino, while its preon hides
inside a $W$ or $Z^0$.
As an example, the decay $\mu^- \rightarrow
\nu_{\mu} + e^- + \bar{\nu}_e$ is
the preon process $\bar{\alpha}
(\bar{\alpha} \bar{\delta}) \rightarrow
\bar{\beta} (\bar{\alpha} \bar{\delta}) +
\beta (\beta \delta) +
\bar{\alpha} (\bar{\beta} \bar{\delta})$,
where the latter two leptons are the decay
products of $W^- = \beta \bar{\alpha}$
(more on vector bosons below).

The observation of three conserved
lepton numbers $(L_e,L_{\mu},L_{\tau})$
in all known weak decays
is a straightforward consequence of
{\it preon stability}. Therefore, this model
is to the best of our knowledge the only one
that explains the empirical lepton-number
conservation from a first principle.
This feature is, however, more or less
accidental for those preon regroupings that
correspond to the known decays, and there is
no general conservation of lepton number in
the model. A simple example is the decay of
the superheavy $\kappa$, {\it e.g.}, through
$\kappa^+ \rightarrow \mu^{+} +
\nu_e + \bar{\nu}_{\tau}$, violating all
three lepton numbers! Such violations are
related to processes that require a break-up
of a spreon, and a regrouping
of its two preons into other spreons.

Other possible lepton-number violating processes
are neutrino {\it oscillations} and {\it decays},
if at least one neutrino is massive.
Indeed, the two superheavy
neutrinos {\it must} decay, an example
being $\bar{\nu}_{\kappa 2} \rightarrow \tau^-
+ \bar{\nu}_{\tau}+ \mu^{+}$.
Such decays relieve us from problems
with cosmic, superheavy and stable neutrinos
surviving from the Big Bang, which would have
given too much cosmic dark matter.

An interesting observation is that the
three neutrinos $\nu_e$,
$\bar{\nu}_{\mu}$ and
$\bar{\nu}_{\kappa 2}$ have
{\it identical} preon content, differing
only in the grouping into spreons:
$\nu_e = \alpha (\beta \delta)$,
$\bar{\nu}_{\mu} = \beta (\alpha \delta)$
and $\bar{\nu}_{\kappa 2} =
\delta (\alpha \beta)$.
This opens up for $\nu_e \leftrightarrow
\bar{\nu}_{\mu}$ oscillations or
{\it electromagnetic}
decays like $\bar{\nu}_{\mu} \rightarrow
\nu_e + \gamma$, which would be the
equivalent of the decay
$\Sigma^0 \rightarrow \Lambda^0 + \gamma$
in the quark model.
It should be noted that the unorthodox
transition between an {\it anti}neutrino and
another neutrino is not
forbidden by helicity conservation, since
a massive neutrino has both left-handed
and right-handed components, and since
the photon carries away helicity.
It should also be observed
that {\it charged} leptons cannot decay
electromagnetically in this model.

Neutrino oscillations can
occur also if there is
a quantum-mechanical mixing of the
${\alpha}$ and ${\delta}$ preons inside spreons,
mixing the $(\alpha \beta)$ with the $(\beta \delta)$
in the quark and lepton wave functions. This is not
forbidden by any known principles, and
would lead to an oscillation between
the $\nu_e$ and the $\nu_{\tau}$.
As will be argued below for quark decays,
there are reasons to believe that such
a mixing occurs on the {\it per mille} level.
This would also mix the $e$ with the $\tau$
by as much, but that would hardly
be observable, since the electron is stable.
(The best way to observe that two
particles are mixtures of the same
wave-function components is to
measure branching ratios when they
decay to one and the same ground state.)

In a future work we will analyse if
any of these oscillations or
electromagnetic decays are in line with the (not fully
consistent) experimental data on neutrino
``transitions'', {\it e.g.}, empirical
limitations on decays \cite{neutrino},
the suppression of
atmospheric muon neutrinos \cite{atmos},
and the recent evidence
\cite{LSND} of a $\nu_{\mu}
\rightarrow \nu_e$ ``oscillation''.

{\it Quark decays} are known to differ
substantially from those of leptons,
in the sense that some quarks clearly mix
quantum-mechanically with each other
when decaying. This 
behavior can be accounted for, but not
explained, by the standard model,
in terms of the CKM matrix
\cite{CKM}. In particular, it is not understood why
there is only one substantial (``Cabibbo'') mixing,
{\it i.e.}, of $d$ and $s$, while the
others are much smaller.

All this can be qualitatively
understood in our model,
but probably not in any other model where
quarks and leptons are related.
The crucial property of quarks is that they are
composed of both preons and antipreons,
giving access to preon-antipreon {\it annihilation}
channels between
certain quarks, or equivalently, giving some
quarks identical net preon flavors.
In particular, 
$d = \beta (\bar{\beta} \bar{\delta)}$ and
$s = \alpha (\bar{\alpha} \bar{\delta})$
have the net flavors of the $\bar{\delta}$
and will mix via
$\alpha \bar{\alpha} \leftrightarrow \beta \bar{\beta}$
(mediated by a composite $Z^0$, or by any
gauge boson).
The effect is weak, due to either the high $Z^0$ mass
or a strong di-preon binding, which suppresses
the $\alpha \bar{\alpha}$ (and $\beta \bar{\beta}$)
wave-function overlaps.
There are similar mixing channels
between $c$ and $t$, and between $b$ and $g$.

The smaller CKM matrix elements cannot be
understood in the same way. Instead,
a weak $\alpha - \delta$ mixing might be
required. It suffices to assume a mixing of only the
deeply bound preons inside a di-preon.
This will mix $u$ with $c$,
$d$ with $b$, and $t$ with $h$.

It is possible that also CP violation
can be explained by such a mixing.
An interesting observation is that the $K^0$
and $\bar{K}^0$ mesons have
identical preon content, although in
different preon/spreon configurations.
This is no longer true if we introduce
a small $\alpha - \delta$ mixing.
If this involves also a quantum-mechanical phase,
it should result in a CP violation in the $K^0$
system. The $D^0$ and $B^0$ mesons do
{\it not} have preon contents
that are identical to those of their antiparticles,
whatever that means for their
possible CP violation.

The {\it vector bosons} come
about as bound preon-antipreon states.
The most likely configurations are
$W^- = \beta \bar{\alpha}$,
$Z^0 = (\alpha \bar{\alpha} - \beta \bar{\beta})/\sqrt{2}$
and $W^+ = \alpha \bar{\beta}$,
in analogy with the spin-$1$ $\rho$ mesons
in the quark model.
There also exist two heavier and
orthogonal combinations ($Z'$ and $Z''$)
of $\alpha \bar{\alpha}$,
$\beta \bar{\beta}$ and
$\delta\bar{\delta}$
(like the $\omega$ and $\phi$
in the quark model). In addition, there
are two neutral states with
$\alpha \bar{\delta}$
and $\delta \bar{\alpha}$,
as well as two charged states of
$\beta \bar{\delta}$ and
$\delta \bar{\beta}$ (like the four
$K^{\ast}$ mesons).
There should be mixings among
these states, especially as they are all heavy
and unstable. One can restrain these
mixings by fits to the known $W$ and $Z$
decay modes, but that would be beyond
the scope of this first qualitative analysis.
It should be added that the lightest $Z'$
could be lighter than the lower
(model-dependent) mass limit
quoted in \cite{rpp}.

In models with composite $W$ and $Z$
one also expects {\it scalar} partners.
The fact that they have not been observed can
be blamed either on high masses or on very weak
couplings to quarks and leptons \cite{preons}.
The latter
seems more realistic since spin-$0$ systems are
normally lighter than those with spin $1$.
The similarity with the mesons in the quark
model, and the considerable mass
difference between the $\pi$ and $\rho$
mesons, make us suspect
that the missing scalars are rather light,
and hence interesting to search for in
existing experimental data. One possibility is
that some of the heavier scalar mesons
are indeed hybrids of quark-antiquark
and preon-antipreon states.

In conclusion, our model provides
qualitative explanations of several phenomena that
are not even addressed in the standard model,
or in other composite models. Most of them
have to do with the (electro)weak interaction, which is
the most troublesome part of the standard model,
with its many {\it ad hoc} parameters
and hypothetical particles and processes.

However, many problems remain to be solved
before it can be considered also a quantitative alternative.
One is the mass-ordering of quarks and leptons,
which does not follow the flavor-$SU(3)$
quark-model recipe of the light baryons.
Obviously, the masses are determined not only
by those of the deeply bound preons,
but also of the exact structure of the composed
objects. It seems as if the ``net'' quantum numbers
are important, so that the colored quarks are
heavier than the color-neutral leptons,
and the normal neutrinos are light because they
carry almost no quantum properties. The
$\bar{\nu}_{\kappa 1}$ and
$\bar{\nu}_{\kappa 2}$ are different
due to the very heavy, ``naked'' $\delta$.

There is also a long way to go before we
understand how preons bind into quarks and
leptons, which is related to the mass problem.
Clearly, the strong binding into di-preons
must be due to attractive
spin forces, {\it e.g.},
color-magnetism in terms of
normal QCD or some even more
fundamental interaction. The binding
of a preon with a di-preon into
quarks is harder to understand. It is not
possible to introduce a new hyper-color,
and require that quarks be
``hyper-confined'' in hyper-color singlets.
Rather it seems as if this binding is due to
a weaker (hyper)color-electric force,
which could be attractive
in both color-singlet (lepton) and color-triplet
(quark) configurations. Then the quarks would
be bound but unconfined, and, in this sense,
less ``fundamental'' than leptons.

Finally, it would be interesting to speculate why nature
has provided two preons ($\alpha$ and $\delta$)
that differ only in mass. Is there some other property that
separates them, and are there more superheavy preons
to be ``discovered'', {\it e.g.}, a heavier version
also of the $\beta$?

Still, as long as preon models have not yet developed into
something as complicated as the standard model
(with the Higgs mechanism), we
consider them worthy of continued
theoretical and experimental efforts.
Many issues, such as the existence of
heavy quarks, leptons and vector bosons,
will hopefully be clarified by the
next generation of high-energy
accelerators.

We acknowledge an illuminating
correspondence with H. Fritzsch, as well as
valuable experimental information
from R. Partridge and G. VanDalen.
This project is supported by the European
Commission under contract CHRX-CT94-0450,
within the network ``The Fundamental Structure of Matter".

\end{document}